\documentclass[letterpaper, 10 pt, conference]{ieeeconf}  

\IEEEoverridecommandlockouts
\title{\LARGE \bf
Provably correct sensor-driven path-following for unicycles using monotonic score functions
}

\author{Benton Clark$^{1}$, Varun Hariprasad$^{2}$, and Hasan A. Poonawala$^{1}$%
\thanks{*This work was not supported by any organization}%
\thanks{$^{1}$Benton Clark and Hasan A. Poonawala are with Faculty of Mechanical and Aerospace Engineering,
        University of Kentucky, Lexington, KY 40506, United States
        {\tt\small \{benton.clark,hasan.poonawala\}@uky.edu}}%
\thanks{$^{2}$ Varun Hariprasad is a student at Paul Laurence Dunbar High School,
1600 Man o' War Boulevard, Lexington, KY 40513, United States
{\tt\small varunhpr5@gmail.com}}%
}

\usepackage[utf8]{inputenc}
\usepackage{subcaption}
\usepackage{graphicx}
\usepackage{amsmath}
\usepackage{amssymb}
\usepackage{todonotes}
\usepackage{leftidx}
\usepackage{tikz}
\usepackage{pgfplots}
\usetikzlibrary{arrows,decorations.pathmorphing,positioning,fit,trees,shapes,shadows,automata,calc,intersections,decorations.markings,pgfplots.fillbetween,patterns} 

\usepackage[normalem]{ulem}
\newcommand{\hlc}[1]{\bgroup\markoverwith
  {\textcolor{#1}{\rule[-1ex]{2pt}{2.5ex}}}\ULon}
\newcommand{\hlctodo}[2]{\todo[color=#1]{#2}\bgroup\markoverwith
  {\textcolor{#1}{\rule[-.5ex]{2pt}{2.5ex}}}\ULon }
\usepackage{hyperref}
\hypersetup{colorlinks=true,     
linkcolor=blue,          
        citecolor=blue,        
          filecolor=magenta,      
           urlcolor=magenta           
}

\newcommand{\ppath}{\mathcal{P}}

\newcommand{\stbsf}{F}
\newcommand{\sebsf}{\Tilde{F}}

\newcommand{\Qone}{\overline{Q}_1}
\newcommand{\Qtwo}{\overline{Q}_2}
\newcommand{\Qthr}{\overline{Q}_3}
\newcommand{\Qfor}{\overline{Q}_4}

\newcommand{\deviation}{d}
\newcommand{\localangle}{\theta}

\newcommand{\R}{\mathbb{R}}
\newcommand{\N}{\mathbb{N}}

\newcommand{\pd}[2]{\frac{\partial #1}{\partial #2}}
\newcommand{\bmat}[1]{\begin{bmatrix}#1\end{bmatrix}}
\newcommand{\bigpar}[1]{\left( #1 \right)}
\newcommand{\sign}[1]{\mathrm{sign} #1 }

\newcommand{\vx}{\vec{x}}

\newcommand{\vxz}{\vec{x}_0}

\newcommand{\vy}{\vec{y}}

\newcommand{\vctrl}{\beta e^{-\alpha \stbsf^2}}
\newcommand{\vctrlsebsf}{\beta e^{-\alpha \sebsf^2}}

\newtheorem{lemma}{Lemma}

\newtheorem{theorem}{Theorem}
\newtheorem{remark}{Remark}
\newtheorem{definition}{Definition}

\begin{document}

\maketitle
\thispagestyle{empty}
\pagestyle{empty}
\begin{abstract}
This paper develops a provably stable sensor-driven controller for path-following applications of robots with unicycle kinematics, one specific class of which is the wheeled mobile robot (WMR). %
The sensor measurement is converted to a scalar value (the score) through some mapping (the score function); the latter may be designed or learned. 
The score is then mapped to forward and angular velocities using a simple rule with three parameters. 
The key contribution is that the correctness of this controller only relies on the score function satisfying monotonicity conditions with respect to the underlying state -- local path coordinates --  instead of achieving specific values at all states. %
The monotonicity conditions may be checked online by moving the WMR, without state estimation, or offline using a generative model of measurements such as in a simulator.
Our approach provides both the practicality of a purely measurement-based control and the correctness of state-based guarantees. 
We demonstrate the effectiveness of this path-following approach on both a simulated and a physical WMR that use a learned score function derived from a binary classifier trained on real depth images. 
\end{abstract}

\section{Introduction}
\label{sec:intro}
Mobile robot navigation is a long-standing topic of research and development in robotics. %
A primary navigation problem is to follow a path~\cite{micaelli1993trajectory,wit1993nonlinear,rio_pathfollowing_1999}, such as defined by lanes on a highway or the walls of a hallway. %
The path is typically assumed free of obstacles, but straying too far from the path may cause collisions. %
Therefore, we would like the robot to stay within some distance from the path due to the possible collisions with objects. 

Traditional approaches to the path-following problem look at designing controllers with provable guarantees of the system stability, assuming direct state measurements are available for feedback~\cite{wit1993nonlinear}. %
The state is typically the distance to a point on the path and the relative heading to the path. %
Measuring these quantities in unstructured environments is difficult or impossible, which limits deployment of mobile robots in such settings. 

In order to avoid direct state measurement, several groups have focused on learning direct sensor-to-actuator feedback controllers~\cite{levine2016end,giusti2015machine}. 
Most of these methods use deep neural networks~\cite{Goodfellow2016DLbook} to map high-dimensional measurements from on-board sensors (e.g.\ optical camera, LiDAR) to control input values, without intermediate state estimation. %
The drawback of these approaches is that they do not provide guarantees on the system stability or convergence to the path for the resulting closed-loop system. %
These methods also require significant data and computational resources~\cite{levine2016end}. 

This lack of guarantees for learned controllers has prompted new controller synthesis and verification methods~\cite{chang2019neural,reedverified,rpm,poonawala2021training}. %
In principle, the methods could enable provably correct sensor-driven control. %
However, they suffer from two drawbacks related to path-following, especially in unstructured environments.
First, the guarantees rely on prior knowledge of a measurement model that predicts measurements that would be obtained around the path. 
This model may be difficult to obtain, and the guarantees are brittle to variation in the real mapping from local path coordinates to sensor measurements that will occur when deployed in changing environments. %
Second, the synthesis and verification of these methods is often computationally expensive, preventing a possible solution where new verified controllers are computed in response to changing environments.

The method in~\cite{poonawala2021training} attempts to overcome the issue of changing environments by using discrete control inputs predicted by a classifier. They argue that the changing measurement model  is equivalent to a variable switching surface in the state space. Using differential inclusion models, they guarantee that deviation from the path are bounded despite uncertainty in the measurement model. This robustness is obtained at the expense of producing a discontinuous switching controller, which is undesirable for real hardware systems. Additionally, the human designer must still choose the number and value of the discrete inputs. 

Ideally, we wish to obtain a sensor-driven path-following controller with guaranteed behavior that does not rely on online state estimation or offline measurement prediction. %

\paragraph*{Contributions} 
This paper proposes a sensor-driven path-following control approach where a high-dimensional sensor measurement is mapped to a scalar value which is in turn mapped to forward and angular velocities of a robot with unicycle kinematics. 
The main contribution is to show that this mapping -- called the Sensor-based Score Function -- from sensor to scalar value must satisfy conditions that do not rely on knowing the measurement obtained in a state \emph{a priori}.  
Additionally, the mapping from scalar value to control inputs is simple with only three parameters that we describe how to choose.
A consequence of this approach is that a measurement model is not necessary when designing the controller, and the guarantees are robust to changes in the measurement model during deployment.
The score function thus enables a connection between pure measurement-based control and rigorous state-based analysis.
We provide results of simulated and physical path-following experiments to demonstrate the benefits of this approach.

\newcommand{\curv}{\rho}
\section{Path-Following Problem}
\label{sec:kinematics}
We consider mobile robots with unicycle kinematics~\cite{malu_kinematics_2014}. Given a one-dimensional curve $\ppath$ in the plane, we can express the kinematics in a local state-space representation known as the ``orthogonal projection''~\cite[p.132]{wit1993nonlinear} of the robot's centroid $P$ about a path $\ppath$. 
The system dynamics consist of the following equations:
\begin{align*}
     \dot s &= v\cos\theta/(1-\curv(s)d), \\ 
     \dot d &= v\sin\theta, \text{ and}\\ 
     \dot \theta &= \omega - v\cos(\theta)\curv(s)/(1-\curv(s)d), 
\end{align*}
where $s$ is the arc length along the curve $\ppath$, 
$\curv(s)$ is the path's curvature at the distance $s$, 
$d$ is the signed distance of the point $P$ from the path, 
$\theta$ is the difference between the heading direction of the WMR and the tangent to the path, 
$v$ is the forward velocity control input, and 
$\omega$ is the angular velocity control input.
The state space values are derived from a Frenet-Serret frame attached along the path $\ppath$ whose normal coincides with the point $P$.
Figure~\ref{fig:ddwmr} provides an illustration of this parameterization.

The \textbf{path-following problem} is to ensure that $d(t) \to 0$ as $t \to \infty$ and that $\dot s(t) > 0$ for all time. 
Note that for paths with time-varying curvature, convergence to the path is impossible without feed-forward control of the path. 
Therefore, our analysis will focus on the case of straight paths.
The convergence guarantees we obtain for the straight-line paths imply bounded deviations for the curved-path case, as argued in~\cite{reedverified}. %

\newcommand{\ddtikz}[3]{
\begin{scope}[xshift=#1cm,yshift=#2cm,rotate=#3,scale = 1.3]
\draw[thick,rounded corners=0.04cm] (-0.3,-0.3) rectangle (0.3,0.3);
\draw[thick,rounded corners=0.04cm,pattern=vertical lines] (-0.5,0.35) rectangle (0.5,0.5);
\draw[thick,rounded corners=0.04cm,pattern=vertical lines] (-0.5,-0.5) rectangle (0.5,-0.35);
\draw[thick] (0.15,0) +(0.25,-0.15) -- +(0.15,-0.1) -- +(0.15,0.1) -- +(0.25,0.15) -- cycle;
\draw[fill = blue,thick] (0,0) circle (0.05) node[left]{P};

\end{scope}
}

\begin{figure}[tb]
\centering
\begin{tikzpicture}[scale=1.0,>=stealth']
\coordinate (orig) at (0,0);
\def\xaxissize{1.5}
\def\yaxissize{1.5}
\def\robotrotate{60}
\def\axisangle{-65}
\ddtikz{0}{0}{60}
\coordinate (fvelbase) at ($(orig)+(10+\robotrotate+\axisangle:0.85*\xaxissize)$);
\draw[->,color=blue,line width = 0.5mm] (fvelbase) -- ($(fvelbase)+(0+\robotrotate:1.0*\xaxissize)$) node[right,blue,text width = 2cm] {forward \\ velocity $v$};
\draw[->,dashed] (fvelbase) -- ($(fvelbase)+(0+\robotrotate:0.75*\xaxissize)$) arc (0+\robotrotate:\robotrotate+\axisangle+90:0.75*\xaxissize);
\node[->,dashed] at ($(fvelbase)+(0+\robotrotate+15:0.9*\xaxissize)$) {$\localangle$}; 
\path[dashed,draw] (fvelbase) -- +(\robotrotate+\axisangle+90:1.0*\xaxissize);
\draw[->,orange,line width = 0.5mm] ($(orig)+(0,0) + (60+\robotrotate:1.0*\xaxissize)$) arc (60+\robotrotate:190+\robotrotate:1.0*\xaxissize) node[right,orange,text width = 2cm] {angular \\ velocity $\omega$};
\coordinate (fsorig) at ($(orig)+(0,0)+ (\robotrotate+\axisangle:2.5*\xaxissize)$);
\draw[dashed] (orig) -- (fsorig);
\path[fill] (fsorig) -- ++(\robotrotate+\axisangle:1.5) coordinate (pathcentre) circle (0.05);
\draw (pathcentre) edge[->] +(\robotrotate+180+\axisangle-30:1.5) +(\robotrotate+180+\axisangle-45:0.75) node[above right, text width = 1.8cm]{path radius $1/\rho$};
\draw[dashed] (pathcentre) +(\robotrotate+180+\axisangle-55:1.5) arc (\robotrotate+180+\axisangle-55:\robotrotate+180+\axisangle+55:1.5);
\path[thick,draw]  (fsorig) to [out =  \robotrotate+\axisangle+90, in = 220] ++(65:1.0*\xaxissize) to [out =  40, in = -40] ++(90:0.2*\xaxissize) node[above] {path};
\path[thick,draw] (fsorig)  to [out = \robotrotate+\axisangle-90,in = 80] ++(-90:0.5*\xaxissize) to [out = -100,in = 30] ++(-145:0.2*\xaxissize);
\draw[very thick,->,red] (fsorig) -- ++(\robotrotate+\axisangle+180:0.95*\xaxissize);
\draw[very thick,->,red] (fsorig) -- ++(\robotrotate+\axisangle+90:0.95*\xaxissize );
\draw[very thick,|<->|,purple] ($(fsorig)+(\robotrotate+\axisangle-90:0.35)$) -- ($(orig)+(\robotrotate+\axisangle-90:0.35)$);
\node[purple] at ($0.5*(fsorig)+0.5*(orig)+(\robotrotate+\axisangle-90:0.6)$) {$\deviation$};
\end{tikzpicture}
\caption{\small A wheeled mobile robot with forward speed $v$, and angular velocity $\omega$. The curved black line represents a local segment of the path, with instantaneous path curvature $\rho$, that the WMR must follow. The local Frenet-Serret frame (red) attached to the path is also shown. The WMR's state consists of the offset $\deviation$ and angle $\localangle$ with respect to the path.}
\label{fig:ddwmr}
\end{figure}
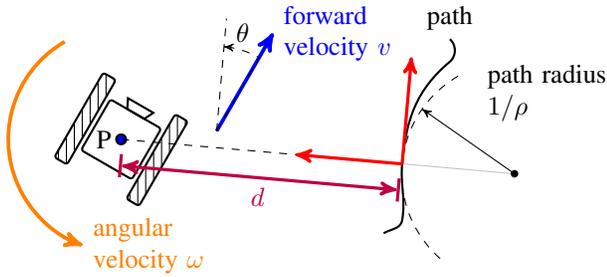

\section{Sensor-driven Path-Following Control}
A sensor-driven controller maps a measurement $\vy \in Y\subset\R^l$ to control inputs $v$ and $\omega$ of the system, where $Y$ is the space of measurements whose dimension is $l \in \N$. %
Instead of designing this controller in an end-to-end manner, we propose mapping the measurement $\vy$ to a scalar value, and then using a simple rule to convert the scalar value into control inputs. 
\subsection{Sensor-Based Score Functions}
\label{sec:sebsf}
We refer to a map $\sebsf: Y\rightarrow\R$ as a \textbf{Sensor-based Score Function} (SeBSF). A SeBSF is an extreme example of dimensionality reduction for high-dimensional sensor measurements, such as optical cameras of LiDAR. 

Sensor-based Score Functions are common, even if not called as such. %
In reinforcement learning where the policy is obtained from a value function over observations~\cite{levine2016end}, the value function is a SeBSF. 
A binary-class support vector machine~\cite{Cortes1995} with input as the sensor image first calculates a score function, whose sign dictates the class. 

As these examples suggest, a distinguishing feature of our work is how we convert the SeBSF into a control law that achieves path-following with guarantees. 

\subsection{Kinematic Control Law}
We propose the following kinematic control law for the mobile robot:
\begin{align}
    \label{eq:omega-ctrl-law} \omega &= \gamma \sebsf, \\
    \label{eq:v-ctrl-law} v &= \vctrlsebsf,
\end{align}
where $\sebsf=\sebsf(\vy)$ is a SeBSF, and $\alpha \ge 0,\beta,\gamma > 0$ are controller parameters.
Parameters $\alpha , \beta,\text{ and }\gamma$ of~\eqref{eq:omega-ctrl-law}-\eqref{eq:v-ctrl-law} correspond to the forward velocity decay rate, the maximum forward velocity, and the angular velocity gain, respectively.

\newcommand{\sensormap}{H}
\subsection{State-Based Score Functions}

To provide guarantees on closed-loop behavior we must convert the sensor-driven control law into a state-based control law. 
To do so, we introduce the notion of a \textbf{State-based Score Function} (StBSF) derived from an SeBSF together with the (unknown) map from state to measurement, which we call the \textbf{sensor map}. %
We define the sensor map formally as follows. 
\begin{definition}
\label{def:sensor-map}
    Let $X\subset\R^2$ be the state space of a WMR as defined in section~\ref{sec:kinematics}.
    Let $Y\subset\R^l$ be the measurement space of an $l$-dimensional sensor.
    Then a sensor map $\sensormap \colon X \to Y$ is a function
    \[ \vy = \sensormap(\vx) , \]
    where $\vx\in X$ and $\vy\in Y$.
\end{definition}
If the sensor map is known \emph{a priori} -- that is, a measurement model is available -- then the conditions we derive can be checked offline. %
For example, we can predict closed-loop behavior in simulation. %
If the map is unknown, the conditions can be checked online without reconstructing the map. %

We now define a State-Based Sensor Function (StBSF).
\begin{definition}
\label{def:stbsf}
    Let $X$, $Y$, $\sensormap$, and $\sebsf$ be a state space, measurement space, sensor map, and SeBSF respectively.
    A State-based Score Function $\stbsf \colon X \rightarrow \R $ is the composition of the SeBSF and sensor map:
    \begin{align}
        \stbsf = \sebsf \circ \sensormap
    \end{align}
\end{definition}
This definition is important as it allows us to analyze the closed-loop behavior due to a sensor-driven control.

\subsection{Conditions On State-Based Score Functions}
In order to guarantee that the path-following is stable and the robot converges to the path, we need that $\stbsf$ be continuously differentiable, and for all states in some region of interest the following conditions hold:
\begin{align}
    F(\vec 0) &= 0 \label{eq:stbsfcondzero}\\
    \pd{F}{\theta} &< 0,\text{ and} \label{eq:stbsfcondpdtheta}\\
    \pd{F}{d} &< 0. \label{eq:stbsfcondpdd}
\end{align}

The important idea is that \textbf{these conditions can be checked online, without estimation of the state}.
We may check~\eqref{eq:stbsfcondzero} by obtaining a measurement when the robot is on the path with its heading aligned with the path tangent. 
We may check~\eqref{eq:stbsfcondpdtheta} by spinning the robot in place while taking measurements. %
We may check~\eqref{eq:stbsfcondpdd} by moving laterally and taking measurements. However, if $| \theta |= \pi/2$ then this last test would be inconclusive.

\section{Closed-Loop Analysis}
\newcommand{\thetainterval}{I_\theta}
\newcommand{\levelset}{c}
\newcommand{\interior}[1]{\text{int}\left({#1}\right)}
\newcommand{\state}{X}
\newcommand{\quadrant}{\overline Q}
This section develops a rigorous proof of the claim that conditions~\eqref{eq:stbsfcondzero}-\eqref{eq:stbsfcondpdd} together with appropriately chosen values of controller parameters ensures that the robot converges to the path. 
For practical selection of the controller parameters, only Remark~\ref{rem:smallbeta} is relevant, with $\alpha\geq 0$. 
\subsection{State-Space Model}
We define the state $\vx(t)$ local path coordinates:
\begin{equation}
    \vx(t) = \bmat{\theta(t) \\ d(t)}.
\end{equation}
Figure~\ref{fig:ddwmr} depicts these quantities. %

We are not concerned with the evolution of $s$, as long as $v >0$ and $| \theta(t)| < \pi/2$. %
A similar choice was made in~\cite{wit1993nonlinear}. 
We also assume there is some maximum allowable deviation from the path $\ppath$ characterized by $d^*$. 
Thus, we consider a compact subset $\state \subset \R^2$ of the state-space  given by 
\begin{equation}
    \state = \{ \vx \in \R^2 \colon | \theta | < \pi/2, | d | < d^*\}
\end{equation}

As mentioned earlier, we analyze the case where the curvature of the path is zero. 
Thus, we focus on the system 
\begin{align}
\label{eq:tdot} \dot\theta(t) &= \omega(t), \text{ and}\\
\label{eq:ddot} \dot d(t) &= v(t)\sin\theta(t).
\end{align}
We denote a solution of~\eqref{eq:tdot}-\eqref{eq:ddot} due to an initial condition $\vxz \in \state$ as $\vx(t;\vxz)$. 

\begin{remark}
\label{rem:nonzerocurvature}
The influence of non-zero curvature is in the form of a disturbance term in the right hand side of~\eqref{eq:tdot}. %
This term is bounded when both the curvature $\rho(s)$ and $d(t)$ are bounded.
The system~\eqref{eq:tdot}-\eqref{eq:ddot} is easily shown to be locally exponentially stable using linearization, implying robustness to bounded disturbances. 
\end{remark}

We now demonstrate that when the state-based score function $\stbsf$ satisfies conditions~\eqref{eq:stbsfcondzero}-\eqref{eq:stbsfcondpdd}, we can find values for parameters $\alpha$, $\beta$ and $\gamma$ such that the origin of the closed-loop system 
\begin{align}
     \dot\theta &= \gamma \stbsf,\label{eq:thetacloop}\\
     \dot d &= \vctrl \sin\theta,\label{eq:dcloop}
\end{align}
is asymptotically stable. 
This closed loop is a result of applying control laws~\eqref{eq:omega-ctrl-law} and~\eqref{eq:v-ctrl-law} to~\eqref{eq:tdot}-\eqref{eq:ddot}, together with converting the SeBSF $\sebsf$ into a StBSF $\stbsf$. 

The proof of asymptotic stability is based on the intuition that the curve $\stbsf(\vx) = 0$ defines a sliding surface. 
If we use a bang-bang control for $\omega$ based on the sign of $\stbsf(\vx)$ and a small enough forward velocity, then the resulting trajectories  would approach this surface rapidly in finite time and then slide along it to the origin. 
Since we use a smooth controller, we instead show convergence to one of a pair of positively invariant cones, and then show that solutions approach the origin once inside these cones. 

Consider the set $F_0$ given by
$$\stbsf_0 = \{ \vx \in \state \colon \stbsf(\vx) = 0\}.$$
We can express the implicit curve $\stbsf_0$ using an explicit function $h \colon \R \to \R$ so that $d = h(\theta)$ for $\vx \in \stbsf_0$. %
\begin{lemma}
\label{lem:explicitrepofF}
Let the StBSF $\stbsf(\vx)$ satisfy conditions~\eqref{eq:stbsfcondzero}-\eqref{eq:stbsfcondpdd}. %
Then, there exists a continuous function $h \colon \R \to \R$ and an interval such that $h(0)=0$ and $\forall \theta$ the following hold:
\begin{enumerate}
    \item $\vx \in F_0 \iff \stbsf(\theta,h(\theta))=0$
    \item $-\infty < h'(\theta) < 0$
    \item $\theta \neq 0 \implies \theta h(\theta) < 0$
\end{enumerate}
\end{lemma}
\begin{proof}
    Since $\pd{F}{d} < 0$ for all $\vx \in \state$, the implicit function theorem guarantees existence of a unique function $h(\theta)$ such that $\stbsf(\theta,h(\theta))=0$. %
    Since $F(0,0) = 0$, we must have $h(0)=0$. 
    The partial derivative of this function is $$h'(\theta) = -\frac{\pd{F}{\theta}}{\pd{F}{d}}.$$ 
    The function $h'(\theta)$ is strictly negative due to~\eqref{eq:stbsfcondpdtheta} and~\eqref{eq:stbsfcondpdd}, and bounded under the assumption that $\stbsf$ is continuously differentiable. 
    Since $h(0) = 0$ and $h'(\theta) < 0$, it is easy to see that $\theta >0 \implies h(\theta)<0$, and  $\theta <0 \implies h(\theta)>0$ . Therefore, $\theta \neq 0 \implies \theta h(\theta) < 0$. 
\end{proof}

Let $\quadrant_i$ be the $\i^{\mathrm{th}}$ quadrant, meaning 
\begin{align}
\Qone &= \{\vx \in \state \colon \theta \geq 0,\, d \geq 0\}\\
\Qtwo &= \{\vx \in \state \colon \theta \leq 0,\, d \geq 0\}\\
\Qthr &= \{\vx \in \state \colon \theta \leq 0,\, d \leq 0\}\\
\Qfor &= \{\vx \in \state \colon \theta \geq 0,\, d \leq 0\}
\end{align}
Given $h$, we can define two lines in the plane, parametrized by their slopes. 
\begin{lemma}
\label{lem:L-bound}
    Consider a curve $h(\theta)$ which is the explicit form of a function $\stbsf$ satisfying conditions~\eqref{eq:stbsfcondzero}-\eqref{eq:stbsfcondpdd}. There exist quantities $L_1$ and $L_2$ where  $0 < L_1, L_2  <\infty$ such that 
    \begin{enumerate}
        \item $-L_1 \theta < h(\theta) <-L_2 \theta$, \text{ and}
        \item $-L_1  < h'(0) < -L_2$.
    \end{enumerate}
\end{lemma}
\begin{proof}
    Since $h(\theta)$ is derived from a state-based score function, by Lemma~\ref{lem:explicitrepofF}, we have $-\infty < -M \leq h'(\theta) \leq -\epsilon < 0$. By simple integration, for any $\theta \in \R$ 
    \begin{align}
        \int_{0}^{\theta} -M d\tau  \leq  \int_{0}^{\theta} h'(\tau) d\tau  \leq \int_{0}^{\theta} -\epsilon d\tau  \\
        \implies  - M \theta \leq  h(\theta)  \leq - \epsilon \theta 
    \end{align}
    Let $0< \varepsilon < \epsilon$. choosing $L_1 = M+\varepsilon$ and $L_2 = \epsilon - \varepsilon$ completes the proof.      
\end{proof}
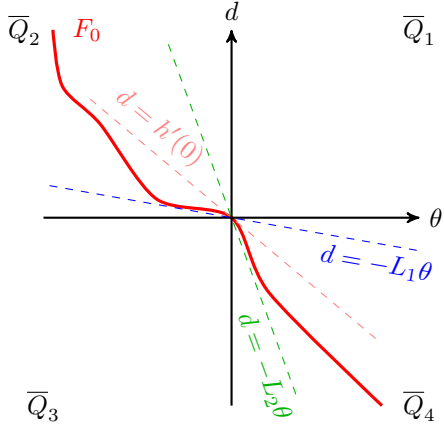
\begin{figure}[tb]
    \centering
    \begin{tikzpicture}[align = center,>=stealth',scale=2.5]
\tikzstyle{myedge}=[shorten >=1pt,auto,semithick]
\tikzstyle{axes}=[->,opacity = 1.0]
\coordinate (A) at (0,0);

\def\cxyline{(0.8,-1) (0.2,-0.4) (-0,0) (-0.4,0.1) (-0.7,0.5)  (-0.9,0.7) (-0.95,1)}

\draw[very thick, red] plot [smooth] coordinates {\cxyline } (-0.9,1) node[right]{$F_0$};
\draw[->,thick] (A) +(-1,0) -- +(1,0) node[right] {$\theta$};
\draw[->,thick] (A) +(0,-1) -- +(0,1) node[above] {$d$};
\draw (1,1) node{$\Qone$} (-1.1,1) node{$\Qtwo$} (-1,-1) node{$\Qthr$} (1,-1) node{$\Qfor$};
\draw[red!50,dashed] (-40:1) --  node[pos=0.8,sloped,above]{$d = h'(0)$} (140:1);
\draw[blue,dashed] (-10:1) --  node[pos=0.1,sloped,below]{$d = -L_1 \theta $} (170:1);
\draw[green!70!black,dashed] (-70:1) --  node[pos=0.1,sloped,below]{$d = -L_2 \theta$} (110:1);
\end{tikzpicture}
    \caption{\small Curves and lines in proof of Lemma~\ref{lem:posinvboundedcones}}
    \label{fig:my_label}
\end{figure}

Lemma~\ref{lem:L-bound} implies that $F_0 \subset \Qtwo \cup \Qfor$. 
The quantities $L_1$ and $L_2$ define two cones $K_1, K_2 \subset \state$ as follows:
\begin{align}
    \label{eq:cone1} K_1 &= \{ \vx \in \state \colon \theta \leq 0,\ -L_1 \theta \leq h(\theta) \leq -L_2 \theta\}, \text{ and}\\
    \label{eq:cone2} K_2 &= \{ \vx \in \state \colon \theta \geq 0,\ -L_1 \theta \leq h(\theta) \leq -L_2 \theta\}.
\end{align}
By construction, $K_1 \subset \Qtwo$ and $K_2 \subset \Qfor$. 

We now show that solutions of~\eqref{eq:thetacloop}-\eqref{eq:dcloop} will enter $K_1 \cup K_2$ in finite time unless they asymptotically approach the origin $\vec 0 \in K_1 \cup K_2$.
\begin{lemma}
\label{lem:coneinfinitetime}
    Consider the set $M = \{ \vx \in \state \colon  \vx \notin K_1 \cup K_2\} $. %
    For each $\vxz \in M$, either $\vx(t;\vxz) \in M\ \forall t\geq 0 \implies \lim_{t \to \infty} \vx(t;\vxz) = 0$ or there exists  $T<\infty$ such that $\vx(T;\vxz) \in K_1 \cup K_2$. 
\end{lemma}
\begin{proof}
    Consider the solution $$\vx(t;\vxz) = (\theta(t;\vxz),d(t;\vxz)),$$ where $\vxz = (\theta_0,d_0) \in M$.
    Let $[0,T(\vxz))$ be the interval over which $\vx(t;\vxz) \in M$ when $\vxz \in M$. 
    
    If $T(\vxz) = \infty$ and $F(\vxz) <0$, then $\dot \theta (t) < 0 \forall t \geq 0$ and $$l = \lim_{t\to \infty} \dot \theta (t) \leq 0.$$ If limit $l=0$, then $\lim_{t\to \infty} F(\vx(t;\vxz)) = 0$, so that the solution asymptotically reaches the line $F_0$. This situation can only occur if $\vx(t;\vxz)$ asymptotically approaches $\vx = 0 \in F_0$. If $\vx(t;\vxz)$ approached any other point in $F_0$, it must have entered $K_1 \cup K_2$ in finite time. Thus, either the limit $l$ is strictly negative, or $T(\vxz) < \infty$. 
    
    If limit $l < 0$, then $\theta(t;\vxz)$ will decrease to $0$ in finite time, say $T'$. 
    If $d(T';\vxz)<0$, then $\vx(t;\vxz)$ must have entered $K_2$ in finite time. 
    If $d(T';\vxz)=0$, then then $\vx(t;\vxz)$ reached the origin in finite time. 
    If $d(T';\vxz)>0$ then $\vx(t;\vxz)$ reaches $\Qtwo$ in finite time. 
    When $\vx(t;\vxz) \in \Qtwo$, then $\theta(t;\vxz) \leq 0$ implying that $\dot d(t) \leq 0$. 
    In turn, $d(t;\vxz)$ decreases, so that $\theta^*(t) = h^{-1}(d(t;\vxz))$ increases. 
    Now, $\theta(t;\vxz)$ will decrease below $\theta^*(t)$ in finite time.
    When $\theta(t;\vxz) = \theta^*$, then $F(\vx(t;\vxz)) = 0$, so that $\vx(t;\vxz)$ must have entered $K_1$ in finite time. 

    An identical argument holds for when $F(\vxz)  > 0$, and is therefore omitted.
\end{proof}
Next, we show that $K_1 \cup K_2$ is positively invariant, and $|d(t)|$ cannot increase when $\vx \in K_1 \cup K_2$. 
\newcommand{\invset}{N}
\begin{lemma}
\label{lem:posinvboundedcones}
    Let \begin{displaymath} \invset_{d^*} = \{ \vx \in \state \colon | d | \leq d^*\}.\end{displaymath}
    Then there exist $\alpha\ge 0$, $\beta > 0$, and $\gamma > 0$ such that the set $\invset_{d^*} \cap (K_1 \cup K_2)$ is positively invariant for all $d^* >0$. 
\end{lemma}
\begin{proof}
Let $\invset_1 = \invset_{d^*} \cap K_1$ and $\invset_2 = \invset_{d^*} \cap K_2$, so that $\invset_1 \cup \invset_2 = \invset_{d^*} \cap (K_1 \cup K_2)$. %
We show that $\invset_1$ is positively invariant, a similar argument holds for $\invset_2$. %
The proof involves showing that the vector field on the boundary is either tangential to or points into the set. 
The set $\invset_1$ is bounded by three lines: $d = d^*$, $d = -L_1 \theta$, and $d = -L_2 \theta$. \newline%
\underline{Line  $d = d^*$:}  By design, $\invset_1 \subset \Qtwo$. 
Therefore, $\forall \vx \in \invset_1$, $\theta \leq 0$. 
By~\eqref{eq:dcloop}, $\dot d = \beta e^{-\alpha F^2} \sin \theta$, so that $\dot d \leq 0$. 
Thus, trajectories cannot leave $\invset_1$ through this line.  \newline 
\underline{Line  $d = -L_2 \theta$:}  We can represent this line by the covector $p_2 = \bmat{L_2 & 1}$, and require that $L_2 \dot \theta + \dot d \leq 0$. We derive
\begin{align}
    &L_2 \dot \theta + \dot d \leq 0\\
    \iff& L_2 \gamma F + \beta e^{-\alpha F^2} \sin \theta  \leq 0\\
    \iff& L_2 \gamma F   \leq \beta e^{-\alpha F^2} (-\sin \theta) \label{eq:L2Nagumo}
\end{align}
For points on the line $d = -L_2 \theta$, $F(\vx)$ is negative, since by Lemma~\ref{lem:L-bound}, $h(\theta) \leq -L_2 \theta$ and $\partial F / \partial d < 0$. Thus, the left hand side of~\eqref{eq:L2Nagumo} is negative. Since $\theta \leq 0$ for all $\vx \in \invset_1$, the right hand side of~\eqref{eq:L2Nagumo} is always non-negative. Therefore, this inequality always holds, and trajectories cannot leave $\invset_1$ through this line. \newline
\underline{Line  $d = -L_1 \theta$:}  Similarly, we can represent this line by the covector $p_1 = \bmat{L_1 & 1}$. 
We require that $L_1 \dot \theta + \dot d \geq 0$ for $\vx$ such that $d = - L_1 \theta$. 
Let $$\stbsf_{L_1}(\theta) = \stbsf(\theta,- L_1 \theta ).$$
We derive
\begin{align}
    &L_1 \dot \theta + \dot d \geq 0 \label{eq:L1Nagumoraw}\\
    \iff& L_1 \gamma \stbsf_{L_1}(\theta) + \beta e^{-\alpha \stbsf_{L_1}(\theta)^2} \sin \theta  \geq 0\\
    \iff& L_1 \gamma \stbsf_{L_1}(\theta)  \geq \beta e^{-\alpha \stbsf_{L_1}(\theta)^2} (-\sin \theta) \label{eq:L1Nagumo}
\end{align}
For $\theta \neq 0$, since $\sin \theta < 0$, we can rewrite~\eqref{eq:L1Nagumo} in the equivalent form
\begin{align}
    \frac{\beta}{\gamma} \leq \frac{L_1  \stbsf_{L_1}(\theta) e^{\alpha \stbsf_{L_1}(\theta)^2}}{-\sin \theta} \label{eq:paramboundL1inv}
\end{align}
For $\theta < 0$, the right hand side of~\eqref{eq:paramboundL1inv} is strictly positive. However, to ensure that~\eqref{eq:L1Nagumo} holds for some $\beta,\gamma >0$, we must confirm that the limit of the right hand side of~\eqref{eq:paramboundL1inv} as $\theta$ approaches zero is strictly positive. 
Taking limits,
\begin{align}
    \Delta = &\lim_{\theta \to 0}  \frac{L_1  \stbsf_{L_1}(\theta) e^{\alpha \stbsf_{L_1}(\theta)^2}}{-\sin \theta} = \frac{0}{0}
\end{align}
Since $\cos 0 \neq 0$, we may use L'Hospital's rule. 
\begin{align}
    \Delta=&\lim_{\theta \to 0}  \frac{L_1  \pd{\stbsf_{L_1}}{\theta} e^{\alpha \stbsf_{L_1}^2} + L_1  \stbsf_{L_1} 2 \alpha \stbsf_{L_1} \pd{\stbsf_{L_1}}{\theta} e^{\alpha \stbsf_{L_1}^2}}{- \cos \theta}\\
    &= \frac{L_1  \pd{\stbsf_{L_1}}{\theta}(0) + 0}{-1} = L_1 \bigpar{-\pd{\stbsf_{L_1}}{\theta}(0) } \label{eq:Delta}
\end{align}
Now, 
\begin{align}
    \pd{\stbsf_{L_1}}{\theta}(0) = \pd{\stbsf}{\theta}(0) + \pd{\stbsf}{d}(0) \cdot (-L_1) \label{eq:FL1prime}
\end{align}
By Lemma~\ref{lem:L-bound}
\begin{align}
    -L_1  <  h'(0)     &\implies -L_1 < -\frac{\pd{F}{\theta}(0)}{\pd{F}{d}(0)} \\
    &\implies \pd{F}{\theta}(0) - L_1 \pd{F}{d}(0) < 0 \label{eq:FL1primeneg}
\end{align}
By~\eqref{eq:FL1prime} and~\eqref{eq:FL1primeneg}, $\Delta > 0$. Thus, if we choose 
\begin{equation}
    \beta < \gamma \Delta, \label{eq:betagammabound}
\end{equation}
then~\eqref{eq:L1Nagumoraw} holds. In turn, trajectories of~\eqref{eq:thetacloop} and~\eqref{eq:dcloop} cannot leave $\invset_1$ through the line $d = -L_1 \theta$. 
Thus, $\invset_1$ is positively invariant for $\beta >0$ and $\gamma >0$ satisfying $\beta < \gamma \Delta$. 
\end{proof}
\begin{remark}
\label{rem:smallbeta}
The quantity $\Delta$ in~\eqref{eq:Delta} may be difficult to estimate. 
Decreasing the ratio $\beta/\gamma$ to a sufficiently small positive value will ensure that~\eqref{eq:betagammabound} is met in practice.    
\end{remark}

\newcommand{\icsetone}{\partial \mathcal N^{+}(\delta)}
\newcommand{\stbsfov}{\overline{\stbsf}}
\newcommand{\neighborhood}[1]{\mathcal N({#1})}
Next, we show that the origin is Lyapunov stable.

\begin{figure}[tb]
    \centering
    \begin{tikzpicture}[align = center,>=stealth',scale=2.5]
\tikzstyle{myedge}=[shorten >=1pt,auto,semithick]
\tikzstyle{axes}=[->,opacity = 1.0]
\coordinate (A) at (0,0);
\def\cxyline{(0.8,-1) (0.2,-0.4) (-0,0) (-0.4,0.1) (-0.7,0.5)  (-0.9,0.7) (-0.95,1)}

\draw[very thick, red,name path=FF] plot [smooth] coordinates {\cxyline } (-0.9,1) node[right]{$F_0$};
\draw[->,thick] (A) +(-1,0) -- +(1,0) node[right] {$\theta$};
\draw[->,thick] (A) +(0,-1) -- +(0,1) node[above] {$d$};
\draw[dashed,name path=deltaminus] (-0.5,-0.9) -- ++(0,1.8);
\draw[dashed,name path=deltaplus] (0.5,-0.9) -- ++(0,1.8);
\draw (-0.5,0) node[below left]{$-\delta$};
\draw (0.5,0) node[below right]{$\delta$};
\path [name intersections={of=FF and deltaplus,by=pplus}];
\path [name intersections={of=FF and deltaminus,by=pminus}];
\draw [thick,green!50,fill opacity=0.2,fill] (pplus) rectangle (pminus);
\path (pminus -| pplus) coordinate (q1);
\path (pminus |- pplus) coordinate (q2);
\path (0,0) coordinate (orig);
\draw[dashed] (pminus) ++(-0.2,0) node[left]{$h(-\delta)$} -- ($(q1)+(0.2,0)$);
\draw[dashed] (pplus) ++(0.2,0) node[right]{$h(\delta)$} -- ($(q2)+(-0.2,0)$);
\draw[very thick, blue] (q1 -| orig)  -- node(mid1)[inner sep=0pt]{} (q1);
\draw[very thick, blue] (q2 -| orig)  -- node(mid2)[inner sep=0pt]{} (q2);
\draw[very thin,->,blue] (mid1) to [bend right] ++(-45:1) node(l1)[right]{$\icsetone$};
\draw[very thin,->,blue] (mid2) to [bend left] (l1.west);
\draw[thick,->,green!70!black] (q2)++(0,0.3) to [bend left] ++(-0.3,-0.1) node[left]{$\neighborhood{\delta}$};

\draw[thick] (q1) to [out=120,in=0] (0,0.5);
\draw[dashed] (-0.8,0.5) -- (1,0.5) node[right]{$h(-\delta)+\beta T(\delta)$};
\draw[thick] (q2) to [out=-20,in=180] (0,-0.95);
\draw[dashed] (-0.3,-0.95) -- (1,-0.95) node[right]{$h(\delta)-\beta T(\delta)$};;

\end{tikzpicture}
    \caption{\small Sets involved in proof of Lemma~\ref{lem:lyap-stab}. Every $\delta >0$ defines a unique $\neighborhood{\delta}$ (light green rectangle). Solutions $\vx(t;\vxz)$ starting inside $\neighborhood{\delta}$ can only exit through $\icsetone$ (thick blue lines). These solutions must reach the $d$-axis, so that $d(t;\vxz)$ remains bounded (see Lemma~\ref{lem:coneinfinitetime}). }
    \label{fig:LSprooffig}
\end{figure}
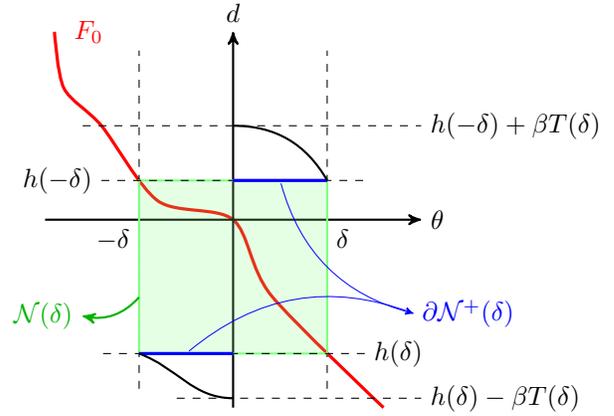

\begin{lemma}
    \label{lem:lyap-stab}
    Let $\beta>0$ and $\gamma>0$ satisfy condition~\eqref{eq:betagammabound}. Then, the origin of~\eqref{eq:thetacloop}-\eqref{eq:dcloop} is Lyapunov stable. 
\end{lemma}
\begin{proof}
    Consider the set $\neighborhood{\delta} \subset \state$ given by
\begin{align}
    \neighborhood{\delta} = \{\vx \in \state \colon |\theta|\leq \delta, h(\delta) \leq d \leq h(-\delta) \}.
\end{align}
This set forms a rectangle (see Figure~\ref{fig:LSprooffig}). Since $h(\theta)$ is monotonic in $\theta$, it is straightforward to show that 
\begin{equation}
    \delta_1 < \delta_2 \iff  \neighborhood{\delta_1} \subset \neighborhood{\delta_2}. \label{eq:monotonicneighborhood}
\end{equation} 
Consider a subset $\icsetone$ of the boundary of $\neighborhood{\delta}$:
\begin{align*}
    \icsetone &= \{ \vx \in  \neighborhood{\delta} \colon \theta \geq 0, d = h(-\delta) \} \cup   \\
     & \qquad \{ \vx \in  \neighborhood{\delta} \colon \theta \leq 0, d = h(\delta) \}.
\end{align*}
    The significance of $\icsetone$ is that any solution $\vx(t;\vxz)$ such that $\vxz \in \neighborhood{\delta}$ can only exit $\neighborhood{\delta}$ through $\icsetone$. To see this, note that for other points on the boundary of $\neighborhood{\delta}$:
    \begin{align*}
        d = h(-\delta), \ \theta <0\implies &\dot d <0, \\
        d = h(\delta), \ \theta >0\implies &\dot d >0, \\
        \theta = \delta \implies F \leq 0 \implies &\dot \theta \leq 0, \text{ or}\\
        \theta = -\delta \implies F \geq 0 \implies &\dot \theta \geq 0.
    \end{align*}
    By Lemma~\ref{lem:coneinfinitetime}, the solutions that leave $\neighborhood{\delta}$ must cross the $\theta = 0$ axis in finite time. 
    
    Consider the set of solutions with initial conditions in $\icsetone$. %
    There exists a maximum value for the time taken to reach the $\theta=0$ axis, otherwise Lemma~\ref{lem:coneinfinitetime} is contradicted. Let this time be $T(\delta)$. Then, because $|\dot d(t)| < \beta$, by integration
    \begin{equation}
       | d(t;\vxz) | \leq \max \bigpar{ h(-\delta) + \beta T(\delta), -h(\delta) + \beta T(\delta) }.
    \end{equation}
    Again, by Lemma~\ref{lem:coneinfinitetime}, solutions $\vx(t;\vxz)$ will reach and stay inside $K_1 \cup K_2$ and $|d(t;\vxz)|$ cannot increase. If solutions cross the line $F_0 \subset K_1 \cup K_2$, then $|\theta(t;\vxz)|$ cannot increase.  Thus, solutions stay within $\neighborhood{\epsilon}$ where $\epsilon = \epsilon(\delta)$ is given by
    \begin{equation*}
        \epsilon(\delta) = \max \bigpar{ h^{-1} \bigpar{ h(-\delta) + \beta T(\delta)}, h^{-1} \bigpar{-h(\delta) + \beta T(\delta)} }.
    \end{equation*}
    Then, all solutions with initial conditions in $\neighborhood{\delta}$ remain inside $\neighborhood{\epsilon(\delta)}$. 
    Due to $h(\theta)$ being monotonic, the dynamics being Lipshitz, and property~\eqref{eq:monotonicneighborhood}, $\epsilon(\delta)$ is invertible. 
    Let $B_\alpha$ represent an open ball of radius $\alpha$. 
    Let $\epsilon'$ be the smallest value such that $ \neighborhood{\epsilon'} \subset B_{\epsilon}$. 
    Let $\delta'$ be the largest value such that $B_{\delta'} \subset \neighborhood{\delta(\epsilon')}$.
    Then, for every $\epsilon>0$, we have found $\delta'$ such that solutions that start in $B_{\delta'}$ remain within $B_{\epsilon}$. 
    Therefore, the origin is Lyapunov stable. 
\end{proof}

Finally, we present the main result.

\begin{theorem}
\label{thm:mainasymptoticstability}
    Let $\stbsf$ satisfy conditions~\eqref{eq:stbsfcondzero}-\eqref{eq:stbsfcondpdd} and the gains $\alpha$, $\beta$ and $\gamma$ be chosen according to Lemma~\ref{lem:posinvboundedcones}. 
    Then, the origin $\vx = 0$ of the closed loop system~\eqref{eq:thetacloop}-\eqref{eq:dcloop} is (locally) asymptotically stable.
\end{theorem}
\begin{proof}
    Lemma~\ref{lem:lyap-stab} shows that under parameter selection such that the ratio $\beta/\gamma$ is sufficiently small, the origin is Lyapunov stable.
    
    Consider the continuously differentiable candidate Lyapunov function $V(\vx) = d^2 / 2$. 
    On the set $K_1 \cup K_2$, $V(0) = 0$ and $V(\vx) > 0$ when $\vx \neq 0$. 
    Its derivative along solutions of~\eqref{eq:thetacloop}-\eqref{eq:dcloop} is
    \begin{align}
        \dot V(t) = d \dot d = d \beta e^{-\alpha F^2} \sin \theta
    \end{align}
    For $\vx \in K_1 \cup K_2$ where $\vx \neq 0$, both $d \neq 0$ and $\theta \neq 0$. 
    Moreover, $\sign(d) \sign(\theta) < 0$. Therefore, $\dot V < 0$. 
    Therefore, solutions $\vx(t;\vxz)$ for $\vxz \in K_1 \cup K_2$ asymptotically reach the origin.
    Thus, if the ratio $\beta/\gamma$ is selected such that Lemmas~\ref{lem:posinvboundedcones} and~\ref{lem:lyap-stab} are satisfied, the previous fact shows that the origin is asymptotically stable.
\end{proof}

\section{Experiments}
Experiments for this work consist of both simulations and physical experiments.
Both sets of experiments consist of a WMR attached with an on-board depth sensor, wherein a SeBSF was used to control the robot.

\subsection{Obtaining a SeBSF}
\label{sec:experimentsebsf}
A support vector machine~\cite{Cortes1995} (SVM) maps vector-valued inputs $y$ to binary class labels. When these labels are $\{+1,-1\}$ then the SVM is given by 
\[ f(\vy) = \sign{(w^T y + c)}, \]
where $w$ and $c$ are the weights and bias terms of the SVM. %
We obtain an SeBSF by training an SVM using labeled data, and then using the argument of the $\sign{}$ function as the sensor-based score: 
\begin{align}
    \sebsf(\vy) = w^T \vy + c, \label{eq:svmscore} 
\end{align}
where $\vy$ represents the vectorized images provided from the depth sensor.

The inputs we use are depth images captured using a Intel RealSense D435i camera affixed to a mobile robot, when the robot is in a corridor. %
The path is defined by the center of the corridor. 
To label these images, we measure the distance $d$ from the center and heading angle $\theta$ relative to the corridor using a ruler and protractor. 
We then label the images according to $\sign{( -\theta - d)}$. %
Note that the label itself does not contain state information, we use the measured state as a guide to labeling. %

\subsection{SeBSF Verification}
In this section, we check whether the SeBSF $\sebsf$ in~\eqref{eq:svmscore} produces a state-based score function that satisfies conditions~\eqref{eq:stbsfcondzero}-\eqref{eq:stbsfcondpdd}. %
The data collected to train the SVM allows us to sample sensor map $\sensormap$ relating an image  $\vy_i$ acquired from the WMR depth sensor with the state $\vx_i$ it was collected in. 

\begin{figure}[t]
\centering
\includegraphics[scale=0.48]{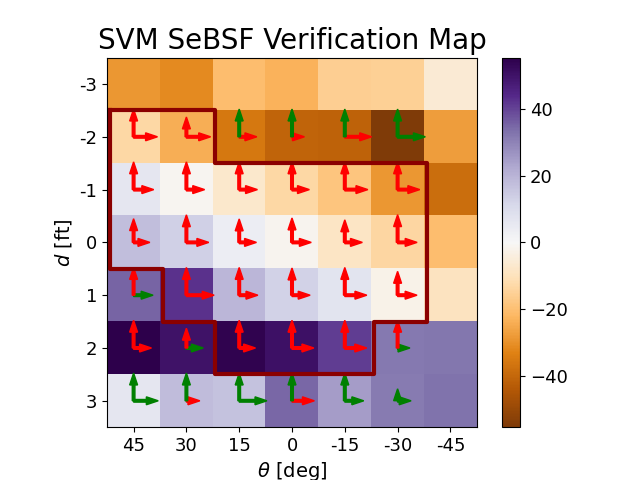}
\caption{\small Plot of the output of the SVM SeBSF verification process.
The vectors at each point represent the local gradient term estimated using finite difference methods.
A red arrow represents $\pd{F}{x_i}<0$, and green opposite.
The maroon curve bounds a region within which both partial derivatives of $\sebsf \circ \sensormap$ are negative.}
\label{fig:sebsf-verification}
\end{figure}

Figure~\ref{fig:sebsf-verification} shows the result of sampling $\sensormap$.
Note that since the SVM was trained as a classifier instead of a regressor, the SVM StBSF $\sebsf \circ H$ is not guaranteed to satisfy conditions~\eqref{eq:stbsfcondzero}-\eqref{eq:stbsfcondpdd}.
However, Figure~\ref{fig:sebsf-verification} shows that within a bounded region of the origin, the StBSF $\sebsf \circ H$ does satisfy~\eqref{eq:stbsfcondpdtheta} and~\eqref{eq:stbsfcondpdd}.
This is the region contained in the maroon boundary shown in Figure~\ref{fig:sebsf-verification}, which contains the origin.
This again highlights the power of the SeBSF --- while the exact score of the SVM is unknown at each point, the partial derivatives are of primary importance for guaranteeing path following.

\subsection{Simulated Experiments}
The simulation experiments were carried out using the AI Habitat simulator~\cite{szot2021habitat,habitat19iccv} in combination with the Habitat-Matterport 3D Dataset~\cite{ramakrishnan2021hm3d} (HM3D) for the navigation mesh map.
The map HM3D-795 was selected for its straight hallway with relatively few obstructions along the length.
The simulator runs with a 100Hz control loop for a fixed interval of time, where the control loop frequency was selected sufficiently fast to estimate a continuous controller on the discrete input WMR used in the simulator.

Within the simulator, a depth sensor was provided to the WMR with characteristics (pixel count, pixel format, etc.) that mimic the Intel RealSense D435i. %
We use the sensor-based score function in~\eqref{eq:svmscore} which is derived from a SVM classifier for \emph{real} images (see Section~\ref{sec:experimentsebsf}) to control the \emph{simulated} robot. %

We simulate 27 trajectories, corresponding to three sets of 9 trajectories corresponding to three different values of ratio $\beta/\gamma$, since this ratio influences closed-loop behavior. %
We choose $\alpha=5\mathrm{e}{-5}$ for all trajectories. %

\begin{figure}[t]
\centering
\includegraphics[width = 0.48\textwidth]{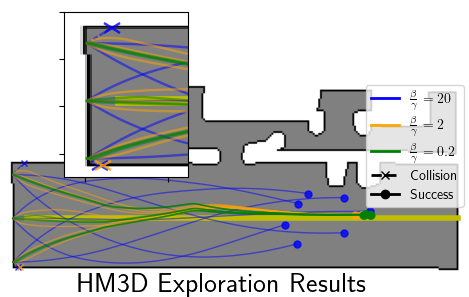}
\caption{\small A top-down view of the trajectories run in the AI Habitat simulator.
The yellow line running the middle of the hallway represents the desired path to follow.
The colors indicate the $\beta/\gamma$ ratio of the trajectories.
Trajectories either end in a crash (marker `X') or don't (marker `O'). 
The inset expands the initial part of the hallway to show the crashes.}
\label{fig:topdown-map}
\end{figure}

\begin{figure}[h]
\centering
\includegraphics[width = 0.5\textwidth]{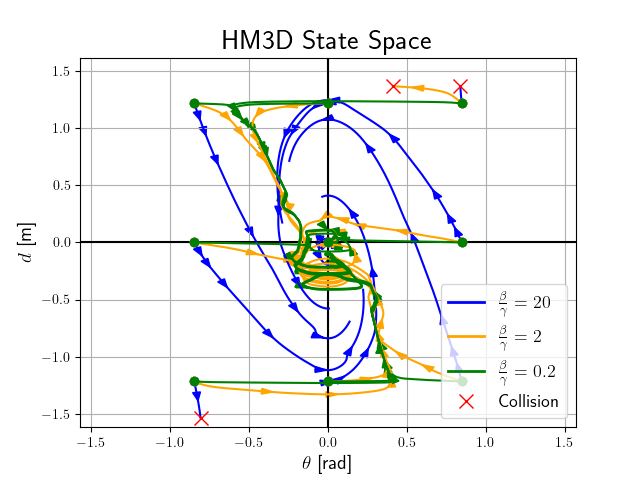}
\caption{\small The state space plots for the trajectories presented in Figure~\ref{fig:topdown-map}.
The trajectories for $\beta/\gamma \in \{0.2,2\}$ appear to trace out a line $F_0$ in $\Qtwo$ and $\Qfor$. The opening in the hallway (see Figure~\ref{fig:topdown-map}) temporarily causes these trajectories to leave $F_0$ in $\Qfor$. For $\beta/\gamma = 20$, there is no invariant cone in $\Qtwo$ or $\Qfor$.
}
\label{fig:ss-trajectories}
\end{figure}

Figure~\ref{fig:topdown-map} shows the trajectory obtained in each of the simulations superimposed over a top-down view of the map. %
Each of the successful trajectories approach the path (yellow line) as expected. %
Figure~\ref{fig:ss-trajectories} shows these trajectories in the  state-space. %
Both figures show that as the ratio $\frac{\beta}{\gamma}$ decreases, the robot converges more quickly to the desired path.
Also, the only set of parameters that did not experience a crash was the lowest value of ratio $\beta/\gamma$, highlighting the importance of this ratio for stability.

We note that there is a minor discrepancy in the path-following exhibited about halfway down the hallway.
This trend was expected as the SVM uses geometric data to predict the score, and the opening in the hallway slightly perturbed the output.
It is useful to see though that the SeBSF is robust to minor variations in the environment.


\begin{figure}[t]
\centering
\includegraphics[scale=0.03]{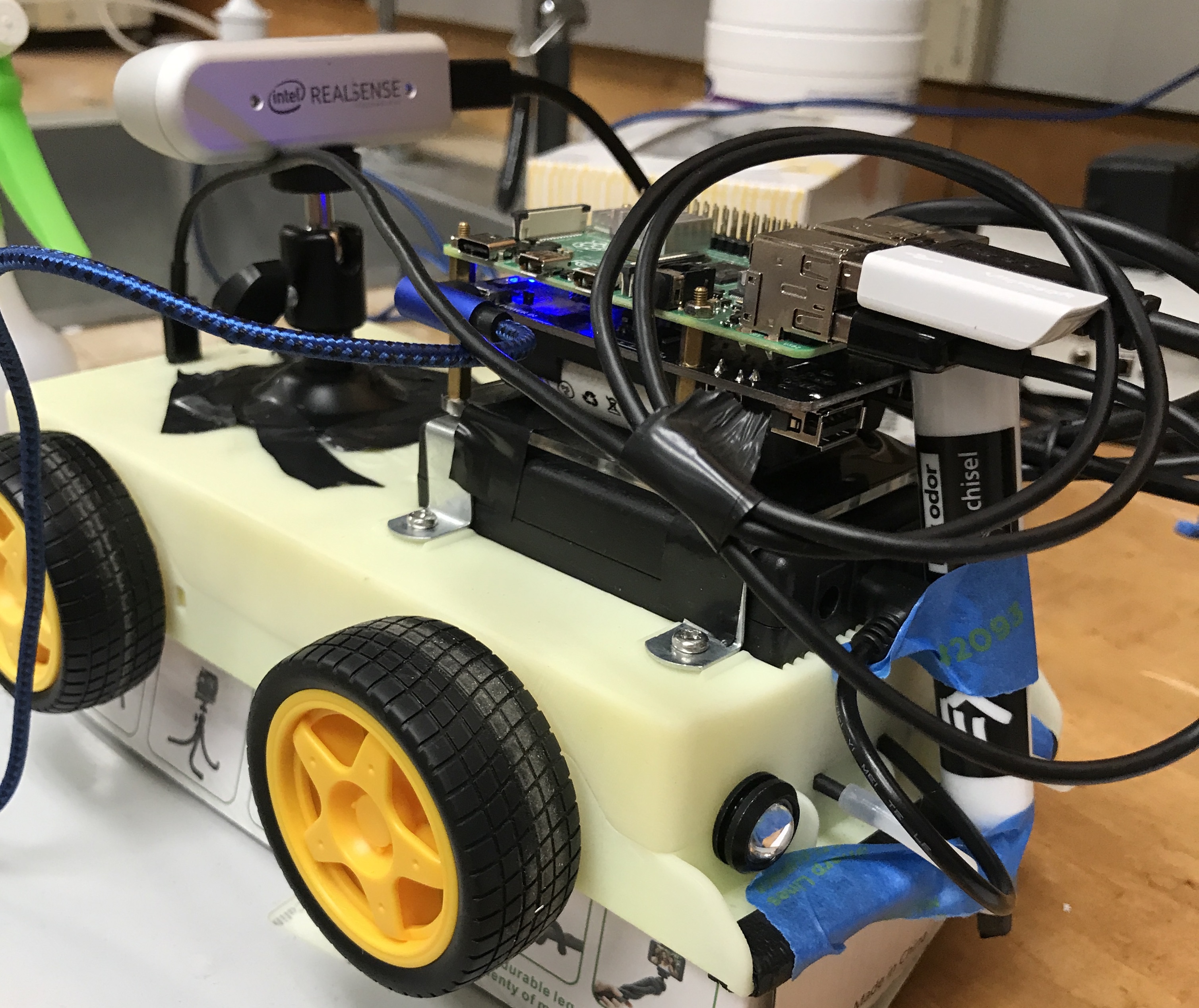}
\caption{\small The platform on which the SeBSF data collection and physical experiments were conducted.}
\label{fig:openbot}
\end{figure}
\subsection{Physical Experiments}
\label{ssec:physical-experiments}
The physical experiments consisted of running the SVM-based SeBSF on a physical mobile robot platform down a hallway.
Figure~\ref{fig:openbot} shows an image of the platform. %
We use a custom-rigged OpenBot~\cite{mueller2021openbot} WMR platform.
The OpenBot platform typically expects a smartphone for the autonomous control.
Instead, we attached an Intel Realsense  D435i camera for imaging, with a RaspberryPi 4 to compute the SVM SeBSF score and send wheel velocity commands to the Arduino driving the motor controllers on the OpenBot.
We extract a depth image with 640x480 pixels in Z16 output format from the RealSense camera at a rate of $15$Hz. 

The experimental setup and data collection was similar to an idea used in~\cite{bakker2010path}.
Each trajectory took place over a 100ft stretch of a straight hallway measuring $8$ft wide.
Three runs were obtained, with a different set of parameters $\beta$ and $\gamma$ chosen for each run.
$\alpha=5\mathrm{e-}5$ was held constant between trials.
A dry-erase marker, visible in Figure~\ref{fig:openbot}, was attached to the back of the OpenBot.
The marker at the back of the car marks the path followed by the robot.
A ruler and protractor were used to measure the distance $d$ and the angle $\theta$ respectively along the trajectory of each run.
The points collected were uniformly spaced by a distance of $2$ft.

\begin{figure}[t]
\centering
\includegraphics[scale=0.55]{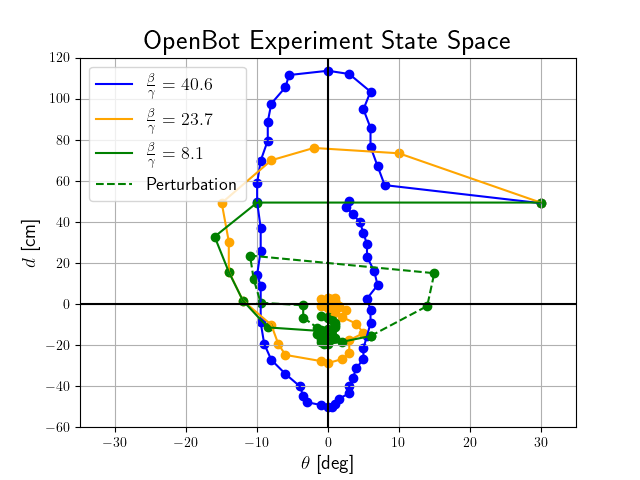}
\caption{A state-space plot of the output from the physical experiments.
Note that the `Perturbation' portion of the trajectory corresponds to the trajectory $\beta/\gamma=8.1$.}
\label{fig:state-space-map}
\end{figure}

Figure~\ref{fig:state-space-map} shows the results of these physical experiments.
We expect that behavior of trajectories to differ from that in simulation due to the low control frequency ($15$Hz).
We varied the ratio $\beta/\gamma$ by keeping $\beta$ fixed and increasing the $\gamma$ (thus decreasing the ratio $\beta/\gamma$).
This approach meant that the WMR rotated more aggressively given the same SeBSF output, which caused issues when running at a relatively low control frequency.
However, the physical experiments do confirm that as the ratio $\beta/\gamma$ decreases, the trajectories converge faster to the desired path.
We also see the trajectories oscillate less as the ratio decreases.

One significant deviation from predicted behavior, labeled as `Perturbation' in Figure~\ref{fig:state-space-map}, occurred when the depth sensor was near a door with a glass pane, thus significantly perturbing the SeBSF output. %
This perturbation only impacted the controller with the largest $\gamma$ value. %
It's impact persists for about $6$ft of the path, then the WMR converges back to the path.

\section{Discussion \& Future Work}
We have both theoretically and empirically demonstrated the path-following guarantees of the continuous controller in combination with a SeBSF score function.
The theory stated that given a SeBSF $\sebsf$ whose StBSF representation $\sebsf\circ H$ satisfies conditions~\eqref{eq:stbsfcondzero}-\eqref{eq:stbsfcondpdd}, and selecting controller gains with a suitably small forward velocity per turning rate, the WMR would converge to the path defined by the sensor reading $\sebsf(\vec 0)=0$.
Our empirical results, both in simulation and on a physical system support this claim.

One point to note however is that the graphs in Figures~\ref{fig:ss-trajectories} and~\ref{fig:state-space-map} did not converge to the origin $\theta=d=0$, but rather converged to a point $(0,d)$ where $d\ne 0$.
We expect that this discrepancy is due to the fact that the SVM did not perfectly achieve the condition $F(\vec 0)=0$. 
However, from Figure~\ref{fig:sebsf-verification}, each of the partial derivatives in that region are negative, and by continuity there must exist a sensor output such that $F(\vx) = 0$. 
We conjecture that the sensor-driven controller is converging to the path defined by the score function being zero, which now differs from our imposed straight-line path. 

Future work will look at a few different ideas.
First, we will extend the following analysis for general curved paths in the plane while maintaining the stabililty guarantees. %
We conjecture that handling curved paths allow us to define paths by the set of robot poses where the \emph{sensor}-based score function is zero, so that condition~\eqref{eq:stbsfcondzero} would automatically be satisfied on such paths. %
Second, we will develop a path following controller for path in 3D, extending the utility of this algorithm to robots such as quadrotors. %
Lastly, we will explore using alternative methods to derive a sensor-based score function, such as using reinforcement learning driven by rewards related to near-collisions.

\bibliographystyle{IEEEtran}
\bibliography{main}
\end{document}